\documentclass[aps,prl,reprint,groupedaddress]{revtex4-1}

\usepackage{graphicx}
\usepackage{amsmath} 
\usepackage{epstopdf}

\begin{document}

%\preprint{}

\title{Path-controlled time reordering of  paired photons in a dressed three-level cascade}


\author{Samir Bounouar*$^1$, Max Strau\ss*$^1$, Alexander Carmele$^2$, Peter Schnauber$^1$, Alexander Thoma$^1$, Manuel Gschrey$^1$, Jan-Hindrik Schulze$^1$, Andr\'{e} Strittmatter$^1$, Sven Rodt$^1$, Andreas Knorr$^2$ and Stephan Reitzenstein$^1$}
%\thanks{}
\affiliation{ \em $^1$Institut f\"ur Festk\"orperphysik, Technische Universit\"at Berlin, 10623 Berlin, Germany\\ $^2$Institut f\"ur Theoretische Physik, Technische Universit\"at Berlin, 10623 Berlin, Germany }


\date{\today}

\begin{abstract}
The two-photon dressing of a "three-level ladder" system, here the ground state, the exciton and the
biexciton of a semiconductor quantum dot, leads to new eigenstates and allows one to manipulate the time ordering of the paired photons without unitary post processing. We show that, after spectral post-selection of the single dressed states, the time ordering of the cascaded photons can be removed or conserved. Our joint experimental and theoretical study demonstrates the high potential of a "ladder" system to be a versatile source of orthogonally polarized, bunched or antibunched pairs of photons.
%[Coupled to a strong laser field, the "ladder" system proves to be a versatile source of coherent heralded 2 or 3 photon states.]

%This photon statistics can be further modified by tuning the strongly coupled laser, allowing the preparation of a particular dressed state and reintroducing a time ordering of the cascaded photons emission. 
\end{abstract}

%\keywords{}

\maketitle

\section{}

%%%%%
%%%
%%%%
%%%%
%%%%
%%%%
%%%%

Resonance fluorescence , i.e. scattering of radiation by atomic systems irradiated by a resonant or quasiresonant laser field, has become a very active field of research in quantum optics in recent years. It has been first predicted \cite{mollow} and demonstrated \cite{schuda,wu} that for a two-level system driven by a high intensity laser, the fluorescence spectrum
is characterized by three components, namely the Mollow triplet. After the demonstration that the overall statistics of the generated photons is subpoissonian \cite{kimble}, the correlation of the different components showed the heralding of the photons coming from the different sidebands \cite{cohen, shrama}. These features were later suc-
cessfully reproduced thanks to the resonant excitation of the excitonic transition in quantum dots (QD) acting as "artifcial atoms" \cite{muller, atature, shih, ulhaq, xu2, He}. Further observations, in the context of coherent control, such as the Rabi oscillations \cite{kamada, Ramsay}, Ramsey interferences \cite{jayakumar}, Autler-Townes splittings \cite{Xu, kabuss} confirmed this "atom-like" behaviour.

The involvement of the biexciton state $|XX\rangle$ of a QD forms with the exciton state $|X\rangle$ and the ground state $|G\rangle$ a Ladder-type system. The latter is highly attractive for the field of quantum communication since it proved to be able to provide on demand polarization-entangled \cite{ondem} and time-bin entangled photons \cite{timebin} and gives access to multiple applications \cite{ Pan, laussy}. Polarization entanglement is usually considered when the excitonic fine structure splitting is smaller than the radiative linewidth of the involved transitions \cite{Young}, making the two recombination paths of $|XX\rangle$ indistinguishable.  This is experimentally very difficult to obtain with standard InGaAs QD, for which even since even uni-axially applied strain tuning is not sufficient to fulfill this condition \cite{jons}. Another proposed alternative is to tune the
dot spectrum to have coincidence of colors across generations, rather than within generations \cite{gershoni}. It relies on the the manipulation of the photons quantum chronology so that the "which-path" information contained by the time ordering of the photons is erased. The underlying unitary transformation is challenging and was originally proposed to be performed with bulky linear optics \cite{ knill}.

In this letter, we present a robust and universally implementable scheme to perform this task through the resonant two-photon driving of a Ladder-type system. Interestingly, this system which was recently uncovered experimentally \cite {hargart, ardelt}, suffered so far from severe experimental restrictions which prevented one from
detecting and analyzing the emitted photons in the time domain, for atoms as well as for QDs. We solve this issue by using determinsitcally fabricated QD-microlenses \cite{gschrey} in combination with high efficiency laser suppression. This allows for the practical use of the generated paired or single photons generated by this excitation scheme. The photons coming from the different dressed states and the perpendicularly polarized remaining bare exciton are correlated. Due the coupling of the field, photons within an emited pair have no ordering in their arrival times, which is in contrast to the natural time ordering of the bare excitonic and biexcitonic photons from the radiative cascade. The non-linear nature of the coherent driving is reflected through two-photon Rabi oscillations observed in the time domain. Detuning of the laser allows for a highly efficient preparation of a particular dressed state and forces the system along a certain path in the dressed radiative cascade. Under these conditions the time ordering of the photons is reestablished. These features are in very good agreements with  theoretically calculated correlations in the dressed states basis.
%The resonant two-photon driving of this Ladder-type system quickly appeared as a very attractive way of coherently controlling such a source of quantum states.

% The detuning of the laser from the virtual state allows to switch the cascaded emission from one path through the dressed states to another, giving the possibility to influence further the time ordering of the detected photons.

Our experiments are carried out on self-assembled  InGaAs/GaAs QDs grown by metal-organic chemical vapor deposition (MOCVD) and  embedded in deterministically fabricated microlenses \cite{gschrey}. These nanophotonic structures allow for an efficient and broadband collection of the excitonic and biexcitonic photons. They also play a positive role on the  focusing of the excitation laser on the target QD, improving the signal to noise ratio of the observed transitions. In comparison to planar structures with usually very poor extraction efficiency this leads to major advantages especially with respect to photon-correlation measurements \cite{ardelt}.
The microlens sample is placed in a Helium flow cryostat and cooled at a temperature of 5 K. By using a cross-polarization configuration, the laser  is efficiently suppressed in order to observe the emission  lines with very reduced laser background.

\begin{figure}[htbp]

\centerline{\includegraphics[width=\linewidth]{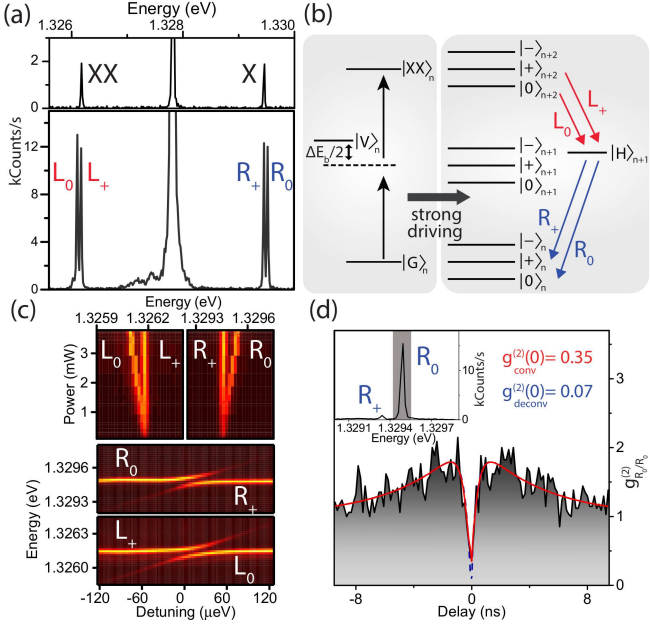}}
\caption{\footnotesize (a) Typical emission spectra of the two-photon resonantly excited QD:  for an excitation power $P_o=100\, \mu\text{W}$ (upper spectrum), for an excitation power of $P=38\, P_o$ (lower spectrum). (b) Left: Scheme of the relevant QD bare excitonic states. Right: Scheme of the dressed states resulting from the two-photon laser coupling to the excitonic states. (c) Upper graph: Spectra of the lines $L_+$, $L_0$, $R_+$ and $R_0$ as a function of the power. Lower graph: spectra of the same lines for different laser detunings to the two-photon resonance. (d) Autocorrelation of the $R_0$ line photons for a laser detuning of $52 \mu eV$. }
\label{fig:gull}

\end{figure}

Fig. 1 (a)  shows  typical spectra of the QD $|G\rangle$-$|X\rangle$-$|XX\rangle$ system being strongly driven by a narrow external cavity tunable laser in resonance to the virtual state of the two-photon resonance for two different excitation powers (upper spectrum for $P_o=100\, \mu$W and lower spectrum for $P=38\, P_o=3.8\, m$W respectively) (i.e. $E_l=E_X-\Delta E_b/2=E_{XX}+\Delta E_b/2$, with $E_l$ the laser energy, $E_X$ the exciton energy and $\Delta E_b$ the biexciton binding energy ). The exciting laser is vertically polarized in order to only involve the vertical excitonic state $|V\rangle$ in the coherent driving and leave the horizontally polarized exciton state $|H\rangle$ unexcited, as can be seen in the underlying level scheme presented in Fig. 1(b). The strong coupling of the laser to the $|G\rangle-|V\rangle$ and $|V\rangle-|B\rangle$ transitions lead to new eigenstates labelled $|+\rangle$, $|-\rangle$ and $|0\rangle$. In particular, the states $|+\rangle$ and $|0\rangle$ can be written:

\begin{align*}
\label{eq:solve}
|0\rangle=\frac{1}{\sqrt{2+(\frac{E_+}{\hbar\Omega})^2}}\Big(|G\rangle+ E_+/(\hbar\Omega)|V\rangle+|XX\rangle\Big)\\
|+\rangle=\frac{1}{\sqrt{2}}(|XX\rangle-|G\rangle)
\end{align*}

$\Omega$ is the (single photon) Rabi frequency and $E_+$ is the eigenenergy of the state $|+\rangle$ (calculated in the supplementary material). The expression of these states show that the state $|+\rangle$ is a simple superposition of $|G\rangle$ and $|XX\rangle$, independant from the Rabi frequency. Its eigenergy is therefore constant with the field strength. On the other hand, the state $|0\rangle$ is a function of the Rabi frequency: its composition and energy change with the excitation power \cite{delvalle}. The resulting emission lines are schematized on Fig. 1 (b). The two lower energy lines labeled $L_0$ and $L_+$ correspond to two transitions from the dressed states of the manifold associated with n+2 interacting photons ($|+\rangle_{n+2}$ and $|0\rangle_{n+2}$) to the bare excitonic state $|H\rangle$. The two lines labeled $R_0$ and $R_+$ on the high energy side correspond to transitions from $|H\rangle$  to the dressed states  of the manifold associated with n interacting photons ($|+\rangle_{n}$ and $|0\rangle_{n}$). Fig. 1 (c) (upper panel) shows the splitting of the dressed states as a function of the excitation power. The increase of the splitting with the excitation power is characteristic of the driving of a ladder system with a resonant laser. As expected from the theory, the energy of the dressed state $|+\rangle$ ($E_+$)  stays unchanged, whereas the state $|0\rangle$ ($E_0)$  is shifted to higher energies \cite{delvalle}.  Interestingly, no  broadening of the dressed state emission lines above the experimental resolution can be observed, which is here a major difference with the phonon-mediated decoherence of the Mollow sidebands \cite{weiler}. This interesting and unexpected feature is most probably related to the fact that in contrast to the standard 2-level system, here the detected transition is not directly driven. Further studies are required to understand the underlying physics in more detail.
% $L_{-}$ on Fig. 1 (b) and 1 (c)

Fig. 1 (c) (lower panel) shows the spectra of the lines around the original $|X\rangle$ and $|XX\rangle$ energies as the laser is swept through the two-photon resonance, at fixed excitation power. A clear anticrossing-like behavior is observed for the two doublets, evidencing the coherent interaction between the laser and the excitonic system.  By laser detunings ($\lvert \Delta_{laser} \rvert> 60$ $\mu$eV), one can prepare a particular dressed state with a great fidelity, and force the cascaded emission of the photon pairs to take a priviledged  path through the dressed states ladder. When the laser is detuned on the higher energy side of the two-photon excitation, the lines $L_+$ and $R_0$ are very intense whereas $L_0$ and $R_+$ are suppressed. The same way, when the laser is detuned towards the low energy side $L_0$ and $R_+$ are the privileged transitions whereas $L_+$ and $R_0$ are suppressed.\\

\begin{figure}[h]

\centerline{\includegraphics[width=\linewidth]{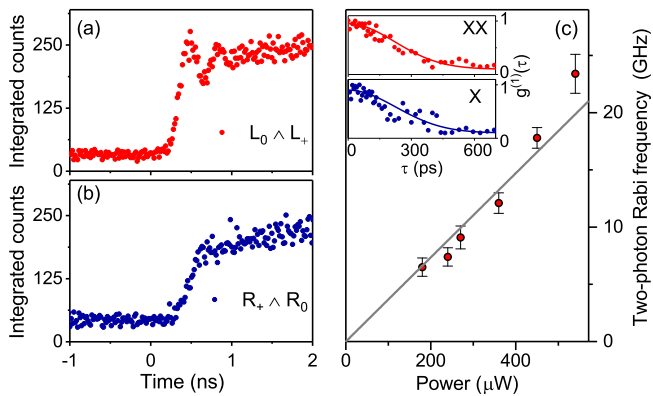}}
\caption{\label{fig1} \footnotesize  (a)  Intensity of the dressed states ($L_++L_0$) after the rise of the 20 ns long laser pulse, (b) Intensity of the excitonic emission ($R_++R_0$) after the rise of the same 20 ns long laser pulse}
\end{figure}

The coherent nature of the two-photon driving is confirmed by the observation of Rabi oscillations in the time domain.
  A two-photon resonant laser pulse at a repetition rate of 17 MHz and 20 ns long (much longer than the lifetime of the transitions $\tau_{(L_0,L_+)}=314$ ps) excites the quantum dot under quasi-CW pumping. The photons emitted by the lines $L_0$ and $L_+$ are detected together by a single photon couting module (SPCM) with 40 ps temporal resolution. The delays between the arrival times of the photons and the trigger (provided by the pulse generator) are measured and stored in a correlation histogram, allowing one to follow the time evolution of the dressed state population under the resonant CW excitation. Fig. 2 (a) shows the resulting measurement. After a non-linear rise of the detection probability, some damped oscillations can be observed, probing the beating of the dressed states between $|XX\rangle$ and $|G\rangle$. These oscillations were until now only probed through picosecond pulsed excitation, as a function of the pulse area \cite{stuffler}. This measurement is here fully accounting for the two-photon resonant driving of the biexciton, where the intermediate exciton state is detuned and therefore adiabatically eliminated leading to an effective two-photon Rabi frequency of $\Omega^2/\Delta E_b$. This measurement provides a dynamical picture of the dressed states population during the coherent driving, whereas in the pulsed configuration one actually measures the occupation of the bare biexciton a long time after the end of the pulse, i.e. after the phonon-mediated relaxation from the dressed states to the bare excitonic states, as the system is not coherently driven any more. 
Fig. 2 (c) shows the extracted Rabi frequencies as a function of the excitation power. Interestingly,  because of the non-linear nature of the two-photon driving, the evolution of the Rabi frequency with respect to the excitation power is linear while, in the case of a two-level system coherently driven by a strongly coupled laser, it scales with square root of the power \cite{wei}. The same measurement was made with the $R_0$ and $R_+$ photons, as shown in Fig. 2 (b). In contrast to the dressed states, no oscillation is here observed: the $|H\rangle$ exciton is not coherently driven and the oscillations probed on the occupation probability of the dressed states are washed out through the radiative relaxation towards $|H\rangle$. These measurements are in good qualitative agreement with the theory presented in the supplementary information \cite{sup} (section V.I).
  Michelson-type interference measurements gives acces to the Fourier transform (defined as $g^{(1)}(t)$) of the spectrum. The first order correlation functions $g^{(1)}(t)$, measured for the ($ L_0+L_+$) lines and for the ($R_0+R_+$) lines are shown in the inset of Fig. 2 (c). They both exhibit a gaussian profil and their fit gives a coherence time of 390 ps $\pm20$ ps for the ($ L_0+L_+$) photons and of 420 ps $\pm$ 30 ps for ($ R_0+R_+$) photons.  The former value is particularly interesting if one considers the very short lifetime of the biexcitonic transitions. These measurements are performed at low power, and the splitting of the dressed states is evaluated to be smaller than $10 \mu eV$. This explains the corresponding oscillations in $g^{(1)}(t)$ with a low bound of the oscillation period of 400 ps are not visible to significant damping at this time-scale.

%scales with $\sqrt{P}$

In order to understand the dynamics of the dressed states and the time ordering of the emitted photons under resonant driving, we correlate the different lines of the emission spectrum with a cross-correlation setup of temporal resolution of 140 ps (full width half-maximum).

% Fig. 3 (a) and (b) show the autocorrelation of the individual lines ($L_0$ and $R_0$). They both exhibit antibunchings. The cross-correlation of  L1 with L2 show %an antibunching, confirming that the dressed states $|+\rangle$ and $|0\rangle$ emit one after the other. The measured characteristic antibunching time ($\tau_p=350 ps$) correspond to %the time nedded by the system to repopulate one dressed state when the other has just emitted. Some remaining backgroung ($g^{(2)}_{L1,L2}(0)=0.25$) can be attributed to some unwanted straylight from the laser and to the contribution of the APD dark counts.

We first checked the autocorrelation of the individual lines $L_0$ and $R_0$, (see inset Fig.1 (d)). As expected, they both exhibit antibunchings. Interestingly, with the laser detuned  (by 52 $\mu eV$) from the two-photon resonance  ($L_+$  suppressed), the correlation of the $L_0$ line is showing a deep antibunching with $g^{(2)}_{L_0}(0)$ value as low as 0.07  limited by non-ideal laser suppression. This indicates that this scheme could be operated as an efficient coherently tunable single photon source, with no need of usually applied strain or temperature-tuning. Here the two-photon driving gives the appealing advantage to offer spectral fine-tuning of coherent single photons emitted far from the laser energy. This could be a substantial improvement for quantum memory schemes requesting the coupling of a quantum dot single photons to an atomic transition \cite{ulrich}, or to the exploration of cavity strong coupling regimes with Mollow triplet sidebands \cite{kim}. The measured characteristic antibunching time ($\tau_p=350\, ps$) corresponds to the time needed by the system to repopulate the next higher two-photon manifold allowing for another two-photon radiative cascade via the non-driven exciton state $|H\rangle$ (see Fig 1 (b)).

In the following, we address the correlations between the single emission lines.
Fig. 3 shows the cross-correlation of the single dressed state lines in two-photon resonance. For the sake of consistency,  negative delays always describe the emission of a "biexcitonic state" ($L_+$ or $L_0$)  after triggering from a "$R_+$ or $R_0$" photon while positive delays correspond to the detection of an "excitonic" photon after a "biexcitonic" photon. Fig. 3 (a) shows the correlation between $L_+$ and $R_+$. Bunchings are observed for positive and negative delays, meaning that the two successive emitted photons are emitted from the same direct radiative cascade without time-ordering. This means that an "excitonic"' photon can also trigger a "biexcitonic"' photon, which is not possible when the radiative cascade is bare. This is due to the fact that this correlated emission of paired photons involves the same dressed state as an initial state and as a final state. The relatively large observed bunching ($g^{(2)}_{L_+,R_+}(0)=4$) indicates that the heralded photons are strongly correlated, presenting obvious advantages in the future realization of Franson-type experiments without the need of tuning the system in a compromise between high count rate and strong correlation, as it is the case for a two-level system \cite{franss}. Fig 3 (b) presents the cross-correlation between the $L_0$ and $R_+$ photons. In this case, photon antibunching is observed for the negative delays. This shows that the original time ordering of the photons is conserved.  This result is explained by the orthogonality of the involved dressed states as described in Eq. 1 and illustrates the versatility of the resonantly-driven radiative cascade: by tuning the spectral detection windows, one has the opportunity to select either time ordered or time unordered paired photons rendering our system an ideal testbed for any time-bin critical experiments.

The laser field is now detuned ($\Delta_{laser}= -63$ $\mu$eV ) from the two-photon resonance. As shown on Fig. 1 (c), under detuned driving, the eigenstates of the system are modified and thus their emission energies. A path though the dressed radiative cascade is priviledged, which has some further consequences on the time ordering of the emitted photons. An example is shown in the inset of Fig. 4. With a detuning $\Delta_{laser}= -63$ $\mu$eV, the lines $L_0$ and $R_+$ become very week whereas the lines $R_0$ and $L_+$ are very bright. The system is actually "forced"' to emit preferentially through the path ($|+\rangle$-$|H\rangle$-$|0\rangle$) and the state $|+\rangle$ is deterministically prepared by the laser field. Fig 4 (a) shows the correlation between the two strong lines of the spectrum $R_0$ and $L_+$. As for the case of strictly resonant exictation, an antibunching can be observed for small negative delays. Despite the fact that the composition of the $|+\rangle$ and $|0\rangle$ states are changed, they remain, as eigenstates, orthogonal to each other.
Fig. 4 (b) shows the correlation between the bright line $R_0$ and the weak line $L_0$. Whereas a bunching can be observed for the positive delays, neither a bunching nor an antibunching can be seen for the negative delays. The detuning of the laser changes the composition of the dressed states, making the emission from the state $|0\rangle$ unprobable. As a consequence, the detection of a $L_0$  photon  triggers with a large probability a $R_0$ photon, whereas the detection probability of a $L_0$ photon after the triggering of a $R_0$ photon becomes so weak that its statistics is poissonian. This effect can be seen as an analogy to the time ordering between Mollow triplet sidebands under detuned excitation \cite{shrama, muller}. Therefore a time ordering is reintroduced between photons which used to be unordered under resonant excitation.

\begin{figure}[h]

\centerline{\includegraphics[trim={0 0.5cm 0 0cm}, width=\linewidth]{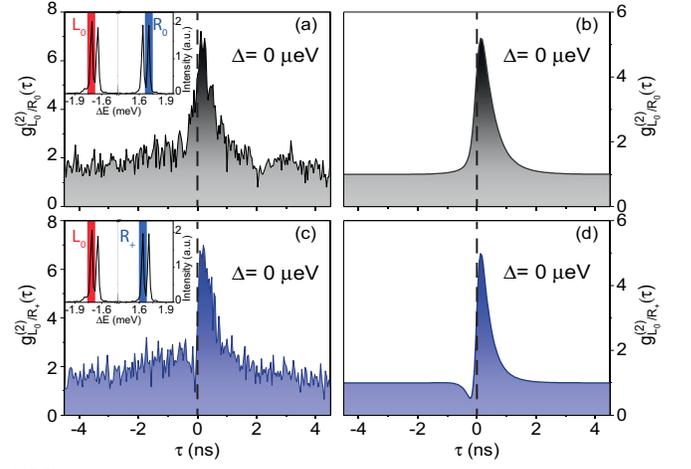}}
\caption{\label{fig1} \footnotesize  Correlations between the two-photon spectrum single lines for the two-photon resonance.  (a) Measured correlation between $L_0$ and $R_0$ for an excitation power of  $P=4.8$ mW: (b) Calculated cross-correlation between $L_0$ and $R_0$ for $\Omega=\Delta E_b/13$, (c) Measured cross-correlation between $L_0$ and $R_+$ for $P=4.8\, mW$, (d) Calculated cross-correlation between $L_+$ and $R_+$ for $\Omega=\Delta E_b/13$. The theoretical curves have been convoluted with the experimental resolution of 140 ps (standard deviation). For the numerical model, measured lifetimes of the exciton and the biexciton ($\tau_{XX}=314$ ps and $\tau_X=742$ ps) were taken as fixed parameters. }

\end{figure}

\begin{figure}[h]

\centerline{\includegraphics[trim={0 0.5cm 0 0cm},width=\linewidth]{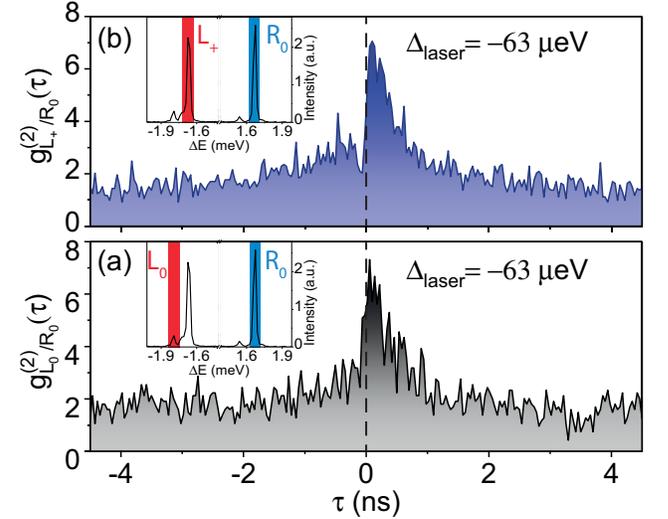}}
\caption{\label{fig1} \footnotesize Correlations between the two-photon spectrum single lines for a laser detuning $\Delta_{laser}= -63$ $\mu$eV  (a) Measured cross-correlation between $L_+$ and $R_0$, (b) Measured cross-correlation between $L_0$ and $R_0$}
\end{figure}

In conclusion, we demonstrated the coherent control over a three-level ladder system and evidenced the two-photon oscillation of its dressed states in the time domain. We show that the strong coupling to the laser field induces a  removal of the photon ordering or its conservation, depending on which dressed states are chosen as initial and final states. Providing that the quantum dot photons can be detected in both polarization components, the time reordering of the photons can be used to complete the "which-path" information erasure \textit{in situ}, without unitary post processing. The large degree of correlation shown by the heralded photons can be used for efficient Franson-type interferometry \cite{franss}.
%This brings the evidences that such resonantly driven $|G\rangle$-$|X\rangle$-$|XX\rangle$  cascades can provide a great variety of photon pairs sources, orthogonally polarized to the laser field, bunched or antibunched.
 It can also be applied  to the generation of entangled photons through a resonant cavity mode under resonant excitation \cite{delvalle}. In the field of quantum communication, it could be directly used in order to enlarge the success probability of quantum teleportation of photonic qubits generated with QDs \cite{shields}.  

%A further study of the photon pairs statistic under slightly detuned excitation promises interesting perspectives. It could lead for instance to a breaking of the symmetry brought by the laser resonance and reintroduce a time ordering of the produced photons. This task would require some additional theoretical efforts on the description of the dressed states of a four-level system under detuned laser driving.
%Such regimes were already explored in the case of the Mollow triplet on atoms and on QD systems \cite{shrama, ulhaq}. This task would require some additional theoretical efforts on the description of the dressed states of a four-level system under detuned laser driving. 

The research leading to these results has received funding from from the European Research Council (ERC) under the European Union’s Seventh Framework ERC Grant Agreement No. 615613 and from the German Research Foundation via Project No. RE2974/4-1 and RE2974/12-1. We also acknowledge support from the Deutsche Forschungsgemeinschaft (DFG) through SFB
787. A. C. gratefully acknowledges support from the SFB 910: "Control of self-organizing nonlinear systems".

*S.B. and M.S. contributed equally to this work.

%\includegraphics[width=90mm]{graph1.pdf}
%\includegraphics[width=90mm]{bunchings1.pdf}

\subsection{}
\subsubsection{}

\end{document}